\documentclass[aps,prb,preprint,showpacs]{revtex4-2}

\usepackage[utf8]{inputenc}
\usepackage[english]{babel}
\usepackage{amssymb,amsmath,mathrsfs}
\usepackage{bm}
\usepackage{graphicx}
\usepackage{ulem}
\usepackage[pdftex,colorlinks=true]{hyperref}

\makeatletter\AtBeginDocument{\let\@elt\relax}\makeatother 

\newcommand{\bra}[1]{\langle #1|}
\newcommand{\ket}[1]{|#1\rangle}

\newcommand{\norm}[1]{\left\lVert#1\right\rVert}

\begin{document}

\title{Spin dynamics from a constrained magnetic Tight-Binding model}

\author{Ramon Cardias}
\email{ramon.cardias@cea.fr}
\affiliation{SPEC, CEA, CNRS, Université Paris-Saclay, CEA Saclay F-91191 Gif-sur-Yvette, FRANCE}
\author{Cyrille Barreteau}
\email{cyrille.barreteau@cea.fr}
\affiliation{SPEC, CEA, CNRS, Université Paris-Saclay, CEA Saclay F-91191 Gif-sur-Yvette, FRANCE}
\author{Pascal Thibaudeau}
\email{pascal.thibaudeau@cea.fr}
\affiliation{CEA, DAM, Le Ripault, BP 16, F-37260, Monts, FRANCE}
\author{Chu Chun Fu}
\email{chuchun.fu@cea.fr}
\affiliation{Université Paris-Saclay, CEA, Service de Recherches de Métallurgie Physique, F-91191, Gif-sur-Yvette, FRANCE}

\begin{abstract}
    A dynamics of the precession of coupled atomic moments in the tight-binding (TB) approximation is presented.
    By implementing an angular penalty functional in the energy that captures the magnetic effective fields self-consistently, the motion of the orientation of the local magnetic moments is observed faster than the variation of their magnitudes. 
    This allows the computation of the effective atomic magnetic fields that are found consistent with the Heisenberg's exchange interaction, by comparison with classical atomistic spin dynamics on Fe, Co and Ni magnetic clusters.      
\end{abstract}

\date{\today}

\maketitle

\section{Introduction}
\label{Introduction}

Nowadays, the coupling between structural and magnetic properties in 3d based magnetic materials plays a key role in the manufacture of high performance spintronics devices~\cite{leeStrongFerroelectricFerromagnet2010}.
Moreover, it is also central in numerous anomalous evolutions of structural parameters~\cite{grimvallLatticeInstabilitiesMetallic2012,*antonangeliAnomalousPressureEvolution2008,*cernyElasticStabilityMagnetic2007,*soderlindCrystalStructureElasticconstant1994} with pressure.
For instance, one of its salient consequence is that the bcc phase of $\alpha$-Fe is stabilized by its magnetic properties~\cite{hsuehMagnetismMechanicalStability2002,*mathonDynamicsMagneticStructural2004,*monzaIronPressureKohn2011}.
Thus, to accurately describe the dynamics of 3d metals and their alloys, a fully coupled spin-lattice dynamics with an {\it ab initio} level of precision is highly desirable.
Unfortunately and despite notable progress~\cite{halilovAdiabaticSpinDynamics1998,maDynamicSimulationStructural2017,*tranchidaMassivelyParallelSymplectic2018}, no such tool is available so far.

However the theory of magnetism is fundamentally a theory of electronic structure.
Antropov {\it et al.} first presented a description of the motion of local magnetic moments in magnetic materials~\cite{antropovInitioSpinDynamics1995,*antropovSpinDynamicsMagnets1996}, in the framework of first-principles methods.
Their idea was motivated by the fact that the interatomic exchange energy among atomic magnetic moments is small compared to intra-atomic exchange and bandwidth energy.
Thus, this adiabatic spin density approximation allows them to treat the angles defining the directions of these magnetic moments as sufficiently slow varying degrees of freedom, to separate them from the individual motion of their underlying electrons, exactly like the nuclear coordinates in the Born-Oppenheimer adiabatic approach to molecular dynamics~\cite{combesBornOppenheimerApproximation1981}.
Moreover, by assuming that the magnetization density in the immediate vicinity of each atom has a uniform orientation, each direction of every magnetic moment can be followed in time according to a precession equation, as it is the case of classical atomistic spin dynamics~\cite{erikssonAtomisticSpinDynamics2017}. 
Consequently, the initial many-electron system is mimicked by this system of classical moments, when the directions and amplitudes are determined self-consistently from the requirement of minimizing a given free energy.
Thus for each moment, the effective field that enters in the precession equation depends only on the variation of the spin-dependant free electronic energy as a functional of the magnetization direction only.
Moreover, by assuming that the relevant electronic correlation hole is essentially in the inner part of each atomic volume, for this type of adiabatic transformation, the longitudinal moment dynamics is nonadiabatic in this approach.
It is governed by individual electronic spin flips like Stoner excitations, which are also fast~\cite{moriyaSpinFluctuationsItinerant2014,*melnikovDynamicSpinFluctuationTheory2018}.
Thus, even if the amplitude of each moment cannot be globally constant in time, for a small temporal excursion fast enough to keep the adiabatic approximation, the longitudinal dynamics can be often neglected.

The paper is organized as followed.
In Sec.~\ref{Methodology}, we review the framework used to derive non-collinear magnetism within the tight-binding (TB) approximation. 
Angular magnetic constraints are imposed by penalty functionals that are solved equally during the self-consistently computation of the electronic structure.
In Sec.~\ref{spin_dynamics}, the derivation of an equation of precession of the local magnetic moments that involves constrained magnetic fields is presented that allows considerations both transverse and longitudinal dampened torques.
The dynamics of various magnetic dimers and trimers of Fe, Co and Ni is studied in details in Secs.~\ref{dimers} and \ref{trimers} to access the validity of the isotropic Heisenberg exchange approximation, that is commonly assumed. 
Lastly, in Sec.~\ref{configdependence} we analyse in depth the example of an Fe dimer exposing the strength of our method as opposed to the limitations introduced by describing this system in the global Heisenberg picture.

\section{Methodology}
\label{Methodology}

When an Hamiltonian $H$ is a functional of the magnetization ${\bm{M}}$, the effective field is nothing else than the functional derivative of $H$ with the respect of the magnetization~\cite{zinn-justinQuantumFieldTheory2002}.
To calculate such an effective field acting on the atomic magnetic moments, the atomistic spin dynamics (ASD) uses a parameterized spin-Hamiltonian, where {\it ab initio} methods calculate it at every self-consistent iteration with various methods.
One of the {\it ab initio} approach consists in the use of constrained density functional theory (cDFT)~\cite{stocksConstrainedLocalMoment1998,*ujfalussyConstrainedDensityFunctional1999,*gebauerMagnonsRealMaterials2000}, where a full accountability of accomplishments of calculations can be found now in many references~\cite{fahnleFastInitioApproach2005,*ujfalussyInitioSpinDynamics2004}. 
The accuracy of the cDFT methods requires an extremely high computational price that scales quickly with the dimension and size of the studied system. 
In contrast, spin-Hamiltonian methods rely on spatial distributions of classical magnetic moments and offer an option with a computational cost tuned by the accuracy and how interatomic exchange parameters are treated.
We offer a method that relies in between, with a lower computational cost compared with the full \textit{ab initio} aspects of the cDFT method without having to rely on a correct description of the parameters inside a spin-Hamiltonian for a given system.

\subsection{Magnetic tight-binding model}
\label{magnetic_TBM}

In this work we have used a magnetic TB model that has been described in a review article~\cite{barreteauEfficientMagneticTightbinding2016} and has been extensively benchmarked and validated in many different magnetic systems of various dimensionalities (bulk, surfaces, interfaces, wires, clusters)~\cite{autesMagnetismIronBulk2006,barreteauMagneticElectronicProperties2012,lelaurentMagnetocrystallineAnisotropyFe2019}, including complex magnetic structures such as spin density wave~\cite{soulairolStructureMagnetismBulk2010} and non collinear configurations \cite{soulairolInterplayMagnetismEnergetics2016}.

It is based on a parametrized $spd$ description of the electronic structure where in practice the parameters of the Hamiltonian are determined by a careful fit of the total energy and band structures obtained from {\sl ab-initio} data over a large range of lattice constants of different crystallographic structures. 
The magnetism is described via a Stoner-like interaction term. The Stoner parameter $I$ of each element being also determined from a comparison to {\sl ab-initio} calculations at several lattice constants. 
This TB model describes the electronic, magnetic and energetic properties with a precision close to Density Functional Theory but at a much smaller computational effort.
 
To avoid a too lengthy derivation, we will present a simplified version of the TB formalism that focuses on the most salient features of the model.
Let us consider a non-magnetic TB Hamiltonian $H^0$ written in a local basis set $\ket{i}$.
The site index $i$ is a composite object that also includes an orbital index reference which can be dropped for simplicity.
$H^0$ is decomposed into onsite energy terms $\varepsilon_i^0=\bra{i}H^0\ket{i}$ and hopping integrals $\beta_{ij}=\bra{i}H^0\ket{j}$.
The eigenfunctions of the system are written as a combination of atomic orbitals $\ket{\alpha}=\sum_i C_i^{\alpha} \ket{i}$ and the density matrix between sites reads $\rho_{ij}=\sum_{\alpha}^{\text{occ}} C_i^{\alpha}C_j^{\alpha \star}$ where the summation runs over the occupied energy levels $\varepsilon_{\alpha}<E_F^0$ where $E_F^0$ is the Fermi level such that $\sum_i \rho_{ii}$ is equal to the total number of electrons $N_e$ of the system.
The total energy of a non-magnetic system is here reduced to the band energy only~\cite{finnisInteratomicForcesCondensed2003}
\begin{align}
    E_{\text{tot}}^0&=\sum_{\alpha}^{\text{occ}} \varepsilon_{\alpha}^0=\mathrm{Tr}(\rho H^0)=\sum_{ij}\rho_{ij}H^0_{ji}\nonumber\\
    &=\sum_{ij}\sum_{\alpha}^\text{occ} C_i^{\alpha}C_j^{\alpha \star}H^0_{ji}.
\end{align}

To this non-magnetic framework, both the magnetic interaction and the local charge neutrality can be added by appropriate constraints, such as the total energy can be written in a formalism where each electronic spins are treated collinear, i.e.
\begin{equation}
    E_{\text{tot}}=E_{\text{tot}}^0 +\sum_i U_i (n_i-n_i^0)^2 -\frac{1}{4}\sum_i I_i m_i^2,
    \label{TB_etot_UI}
\end{equation}
where $n_i=\rho_{ii}=n_{i \uparrow}+n_{i \downarrow}$ and $m_i=n_{i \uparrow}-n_{i \downarrow}$ are respectively the charge and magnetization of site $i$, whereas $I_i$ is the Stoner parameter and $U_i$ a large positive quantity.
By minimizing Eq.\eqref{TB_etot_UI} with respect to the normalized coefficient $C_i^{\alpha}$, with the condition $\sum_i (C_i^{\alpha})^2=1$, this leads to a Schrödinger equation for a renormalized Hamiltonian $H_{\sigma}$ for $\uparrow$ or $\downarrow$ spins separately.
This Hamiltonian simply reads as
\begin{equation}
   H_{\sigma}=H^0 + \sum_{i}  \ket{i}\left(U_i (n_i-n_i^0) -\frac{1}{2} I_i m_i \sigma\right) \bra{i},
\end{equation}
where $\sigma=\pm 1$ is the spin $\uparrow$ or $\downarrow$. 
In this Stoner picture only the onsite terms $\varepsilon^0_i \rightarrow \varepsilon^0_i+(U_i (n_i-n_i^0) -\frac{1}{2} I_i m_i \sigma )$ are affected by both the local charge neutrality and magnetism. 

The generalization to non-collinear magnetism is straightforward.
First the previous expressions is extended to spin-orbitals with spin-dependent coefficients $(C_{i \uparrow}, C_{i \downarrow})$ on each site.
Then an onsite density matrix $\tilde{\rho}_i$ is manipulated as a $2\times2$ matrix with components $\rho_{i}^{\sigma \sigma'}=\sum_{\alpha}^{\text{occ}}C_{i\sigma}^{\alpha}C_{i\sigma'}^{\alpha \star}$, in order to write it more conveniently as $\tilde{\rho}_i=\frac{1}{2}n_i \sigma_0+\frac{1}{2}\bm{m}_i\cdot\bm{\sigma}$, where $\sigma_0$ is the identity matrix $\equiv\mathbb{I}$ and  ${\bm\sigma}=(\sigma_x,\sigma_y,\sigma_z)$ is a vector of Pauli matrices, ${\bm{m}}_i=\text{Tr} (\tilde{\rho}_i {\bm{\sigma}})$.
As a consequence, the Hamiltonian $H$ then reads as
\begin{equation}
   H=H_n\sigma_0+\bm{H}_m.\bm{\sigma},
\end{equation}
where the components of the vector Hamiltonian $\bm{H}=(H_n,\bm{H}_m)$ are
\begin{align}
   H_n&=\sum_i\left(\epsilon_i^0+U_i(n_i-n_i^0)\right)\ket{i}\bra{i}+\sum_{ij}\beta_{ij}\ket{i}\bra{j},\label{H0}\\
   \bm{H}_m&=-\frac{1}{2}\sum_{i}  \bm{\Delta}_i\ket{i}\bra{i}.\label{Hvec}
\end{align}
with $\bm{\Delta}_i=I_i \bm{m}_i$.
When the total energy of the system is written as the sum of the occupied eigenvalues (band energy term) of the renormalized Hamiltonian, one has to take into account the so-called double counting terms
\begin{equation}
    E_{\text{tot}}=\sum_{ \alpha }^{\text{occ}} \varepsilon_{\alpha}-\frac{1}{2}\sum_i U_i ((n_i)^2-(n_i^0)^2) +\frac{1}{4}\sum_i I_i \norm{\bm{m}_i}^2,
\end{equation}
where $\varepsilon_{\alpha}$ are the eigenvalues of the renormalized Hamiltonian. 

\subsection{Magnetic constraints in TB\label{magnetic_constaints_TB}}

When dealing with magnetic systems it is often interesting to be able to explore the energetics of various magnetic configurations.
This can achieved by trying several starting magnetic configurations but remains a relatively limited strategy since this produces few self-consistent solutions to compare with.
It can be very interesting to consider the situation where magnetic constraints are imposed on any given atom $i$ of the system.
Appendix~\ref{fixed_spin} summarizes the fixed spin method that is limited to collinear magnetism.
However, among all the practical methods of optimization under constraints~\cite{fletcherPracticalMethodsOptimization2008}, the penalty method is a very handy way to proceed.

This consists to supplement the total energy with a penalty term in a similar way that has been done for the local charge neutrality constraint.
There exists many possible ways to impose constraints on a magnetic system~\cite{dederichsGroundStatesConstrained1984,gebauerMagnonsRealMaterials2000,maConstrainedDensityFunctional2015}, 
which have been carefully reported in the reference~\cite{kadukConstrainedDensityFunctional2012}.

There also exists various types of penalty functional depending on the quantity to impose.
One can impose a given moment $\bm{m}_i^{\text pen}$ on a given atomic site $i$ as presented in appendix \ref{pen_fixed_spin} but it is also possible to constrain only the polar angle $\theta_i$ between the atomic moments of atom $i$ and the $z$-axis, a penalty functional of the form $\lambda (\theta_i-\theta_i^{\text pen})^2$ can be considered.
An equivalent expression can apply to the azimuthal angle $\phi_i$ too.
To constraint simultaneously both angles, we could simply add these two functionals.
However as reported by Ma and Dudarev~\cite{maConstrainedDensityFunctional2015}, a combined angular penalty functional can be constructed, based on the dot product of $\bm{m}_i$ and $\bm{e}_i^{\text pen}$, here considered as a unit vector of given spherical angles $(\theta_i^{\text pen},\phi_i^{\text pen})$.
This penalty function reads $E^{\text pen}_i=\lambda (\norm{\bm{m}_i}-\bm{e}_i^{\text pen} \cdot \bm{m}_i)$, and leaves the norm of the magnetization $\norm{\bm{m}_i}$ free to vary while the direction of the magnetic moment is constraint to be the direction of $\bm{e}_i^{\text pen}$.
Consequently, this introduces a renormalization of the on-site terms of the TB Hamiltonian of the form $-\bm{B}_i^{\text pen}\cdot \bm{\sigma}$ with $\bm{B}_i^{\text pen}=-\lambda(\bm{e}_i-\bm{e}_i^{\text pen})$, where $\bm{m}_i=\norm{\bm{m}_i}\bm{e}_i $. Therefore the on-site term $\bm{\Delta}_i$ of the magnetic Hamiltonian $ \bm{H}_m$ (see Eq.~(\ref{Hvec})) reads:
\begin{equation} 
    \bm{\Delta}_i=I_i\bm{m}_i+2\bm{B}_i^{\text pen} \label{Eq.Delta}
\end{equation}
This is exactly Eq.~(1.9) of Ref.~\onlinecite{smallCoupleMethodCalculating1984}.
The spin splitting field $\bm{\Delta}_i$ is the sum of the Stoner-like exchange field $I_i\bm{m}_i$ and the penalization field. 
This penalty scheme has many specific properties.
For example by noting that $-\bm{B}_i^{\text pen}\cdot \bm{m}_i=E^{\text pen}_i$, it can be shown that there are no double counting terms associated to the the renormalization.
Consequently the total energy can we written as in Eq.~\eqref{eq:double-counting} but without the last term.
Moreover when $\lambda \to \infty$, $\bm{e}_i \approx \bm{e}_i^{\text pen}$ and $\bm{B}_i^{\text pen} \cdot \bm{m}_i=0$ and the penalization field becomes perpendicular to the local magnetization. 

To be more specific, let us now consider the variation of the total energy with respect to the polar and azimuthal angles.
By considering a variation of angle $d\theta$ on site $i$ and by using the Force Theorem, it is straightforward to show that $dE=-\frac{d\bm{B}_i^{\text pen}}{d\theta}\cdot\bm{m}_i d\theta= -\norm{\bm{m}_i}\frac{d\bm{B}_i^{\text pen}}{d\theta_i}\cdot\bm{e}_i d\theta_i $.
Now by taking the derivative of $\bm{B}_i^{\text pen} \cdot \bm{e}_i=0$, and by noting that $\frac{d\bm{e}}{d\theta}=\bm{e}_{\theta}$, we find a relationship between the polar angle variation of the energy, which is the effective field up to a sign, and the penalty field
\begin{equation}
   \frac{1}{\norm{\bm{m}_i}}\frac{\partial E}{\partial \theta_i}=\bm{B}_i^{\text pen}\cdot\bm{e}_{i, \theta} =\bm{B}_{i,\theta}^{\text pen} ,
\end{equation}
and similarly with the azimuthal angle variation of the energy 
\begin{equation}
     \frac{1}{\norm{\bm{m}_i}} \frac{1}{\sin \theta_i}\frac{\partial E}{\partial \phi_i}=\bm{B}_i^{\text pen}\cdot\bm{e}_{i,\phi}= \bm{B}_{i,\phi}^{\text pen}.
\end{equation}
Or in a more compact formulation 
\begin{equation}
\bm{B}_{i}^{\text {pen}}=\frac{\partial E}{\partial \bm{m}_i}=\frac{1}{\norm{\bm{m}_i}}\frac{\partial E}{\partial \bm{e}_i}.
\end{equation}
Thanks to these penalty functionals, it becomes possible to target any local arbitrary magnetic configuration to find the corresponding local effective field, which is an extremely useful technique to explore the magnetic energy landscape.
It is also possible to assign $\lambda$ as a site-dependent parameter, by setting it to zero to constraint some atoms and let the others to adapt, during the self-consistency cycles. 

In the following section we will use the penalty formalism to map the TB model onto an Heisenberg Hamiltonian and to derive a spin dynamics equation of motion that directly use the penalty field hence derived.

\subsection{Exchange parameters in TB}
\label{exchange_parameters}

In this section the general features to map the total energy of an electronic structure method onto a classical Heisenberg model is presented, that describes a system of atomic spin, characterized by local magnetic moments $\bm{m}_i$ at site $i$ interacting via bare isotropic interactions $J^0_{ij}$:
\begin{equation}
    \begin{aligned}
    E_{\textrm{Heis}}&=-\frac{1}{2}\sum_{i \ne j} J^0_{ij} \bm{m}_i \cdot \bm{m}_j,\\
    &=-\frac{1}{2}\sum_{i \ne j} J^0_{ij}\norm{\bm{m}_i}\norm{\bm{m}_j} \bm{e}_i \cdot \bm{e}_j,\\
    &=-\frac{1}{2}\sum_{i \ne j} J_{ij} \bm{e}_i \cdot \bm{e}_j,
    \end{aligned}
    \label{eq1}
\end{equation}
Within this approach the amplitude of the magnetization $\norm{\bm{m}_i}$ of site $i$ can be incorporated effectively into the bare exchange interaction to produce a dressed exchange interaction, once assumed that the $\norm{\bm{m}_i}$ become independent of the magnetic configuration.
This assumption seems rather drastic but in many magnetic systems, where the magnetic moments are not so dependent on the magnetic configuration or for small rotations around a given angle, which is the case treated here.
By keeping this assumption in mind, we can safely dropped the dressed reference. 

However in systems that break globally the symmetry of space rotation (particularly of nanometer size), this fails and the classical Heisenberg model is only valid for a limited range around a given magnetic stable (or metastable) configuration $\cal C$, that preserves the invariance by point rotation only locally.
In such systems the Heisenberg model can only be used to explore the dynamic around configuration $\cal C$, that does not alter substantially the invariance by point rotation, that are often found for low temperatures.
Consequently for higher temperatures or space transitions that reduce the point symmetry, the $J_{ij}$'s become usually very sensitive to the structural parameters such as the interatomic distances and local environments, preventing their transferability to various atomic structures.
This point is well illustrated in Appendix~\ref{calculated_exchange}.

Since numerical implementations of the Heisenberg model are by far simpler than electronic structure approaches, it is tempting to extract the desired exchange parameters $J_{ij}$ from electronic structure calculations.
To do so, several methods have been reported in the literature.
i) The simplest method is based on a fit of the total energy obtained by multiple magnetic collinear configurations, which do not necessitate any non-collinear numerical implementations neither penalty constraints~\cite{phillipsBlackboxComputationMagnetic2013,*vaclavkovaMagnetoelasticInteractionTwodimensional2020}.
ii) Another approach consists in performing finite difference calculations of the total energy between various magnetic non-collinear configurations~\cite{sandratskiiEnergyBandStructure1986,*sandratskiiNoncollinearMagnetismItinerantelectron1998,*grotheerInitioTreatmentNoncollinear2000,*grotheerFastInitioMethods2001}, which can enlarged significantly the space of the magnetic configurations to span.
In addition by varying the relative angle between the magnetic sites, it is possible to test the range of validity of the Heisenberg picture~\cite{rosengaardFinitetemperatureStudyItinerant1997,brinkerChiralBiquadraticPair2019}.
iii) Based on this finite difference picture, in a seminal work Liechtenstein {\sl et al} derived an explicit expression of the exchange parameters, based on second order variation of the band energy term relying on the magnetic Force Theorem and Green's function formalism~\cite{liechtensteinExchangeInteractionsSpinwave1984,*liechtensteinLocalSpinDensity1987}.
The latter one has shown big success in predicting various magnetic properties such as magnon excitation, critical temperature and also used to perform dynamical calculation of magnetic moments~\cite{pajdaInitioCalculationsExchange2001}.
In this work, we have used the approach ii), where we rotated one magnetic moment of an angle $\theta$ and developed an equation for $E(\theta)$ for each case, e.g. dimers (Sec. \ref{dimers}) and trimers (Sec. \ref{trimers}).
We have found that the energy curve between the TB model and the Heisenberg model agree quite well, which leads to a good agreement between the spin dynamics of the two different methods, shown later in Secs. \ref{dimers} and \ref{trimers}.

Details of the derived expression for both cases and the fitting of the energies to find the respective exchange coupling parameter $J_{ij}$ for each case is explored in more details at the Appendix~\ref{calculated_exchange}.

\subsection{Spin-dynamics in TB}
\label{spin_dynamics}
 
The change in direction of each of the local magnetic moments ${\bm m}_i=\text{Tr} (\tilde{\rho}_i {\bm{\sigma}})$ with time is given by the transverse torque of this moment only with the effective pulsation, which is in return precisely $\bm{B}_i^{\text{eff}}\equiv -\bm{B}_i^{\text{pen}}=-\frac{\partial E}{\partial \bm{m}_i}$,
\begin{equation}
    \label{precession_eq}
   \frac{d\bm{m}_i}{dt}= \bm{m}_i  \times  \frac{\bm{B}_i^{\text{eff}}}{\hbar} =\frac{\bm{B}_i^{\text{pen}}}{\hbar} \times \bm{m}_i 
\end{equation}
Because $\bm{B}_i^{\text{eff}}$ is constructed orthogonal to $\bm{m}_i$, $\bm{B}_i^{\text{eff}}$ is itself a cross product of a functional of $\bm{m}_i$, by $\bm{m}_i$.
Eq.~\eqref{precession_eq} is nothing else than the Larmor's precession equation, which is itself a non-relativistic limit of a more complex motion of spinning particles in a co-moving frame \cite{bargmannPrecessionPolarizationParticles1959}.
 
In practice, TB SCF calculations are first performed without any constraint to identify the stable magnetic (or metastable) states ${\bm{m}^{eq}_i}$.
Such a magnetic state is not necessarily unique and the process has to be repeated in frustrated systems that produce degenerate states.
However this process can be systematized by considering methods for finding minimum energy paths of transitions in magnetic systems~\cite{bessarabMethodFindingMechanism2015,*ivanovEfficientOptimizationMethod2020}.
Moreover if a precession around the equilibrium magnetization is considered, the longitudinal term vanishes because $\bm{B}_i^{\text{eff}}$ is constructed orthogonal to ${\bm{m}_i}$.
Then a given spin direction $\bm{m}_i(0)$ is chosen in the neighborhood of this equilibrium state and a constrained SCF calculation is performed according to the chosen penalty method described above, to get the local effective field.
Thus, a spin dynamics is produced by solving Eq.\eqref{precession_eq} in time by using an explicit solver.
In this case, each local moment may have different starting amplitude, that remains constant over time and their motion evolve on local spheres, according to the Rodrigues' rotation formula, that is presented in Appendix~\ref{spindynamics_dimers}.
The procedure is repeated for each time step of the spin dynamics.

\section{Spin dynamics of magnetic clusters}
\label{clusters}

In this section, we study the dynamics of the magnetic moments under two different scenarios: using an "in house" atomic spin dynamics (ASD) as implemented in Ref.~\cite{beaujouanAnisotropicMagneticMolecular2012} based on an Heisenberg Hamiltonian and the tight-binding spin dynamics (TBSD) method described in the previous Sec.~\ref{Methodology}.
This is applied for the most simple cases, i.e. dimers and equilateral triangle trimers of equivalent atoms for which the corresponding effective exchange interaction $J$ is obtained from our TB model and then used in the ASD for comparison with TBSD. Note that since in the ASD code the dynamics is expressed in terms of unit vectors and the effective field is written as $-\frac{\partial E}{\partial \bm{e}_i}$ (with no $\norm{m_i}$ factor) we have used in the TBSD an effective field  given by $-\norm{m_i}\bm{B}_i^{\text{pen}}$. 

We would like to highlight that Ref.~\cite{streibEquationMotionConstraining2020} have explored aspects of the results presented in this paper, in parallel.
Most of their efforts was to verify if the effective field is exactly the negative of the constraining field, which acts as a Lagrange multiplier to stabilize an out-of-equilibrium, noncollinear magnetic configuration, a point raised in Ref.~\onlinecite{dederichsGroundStatesConstrained1984}.
However, the quality of the derived effective field by constrained method is very sensitive to the numerical limit of the Lagrange multiplier, a point we have carefully monitored.
It is noteworthy to say that our results are complementary and do not overlap in any way, specially in the spin-dynamics aspect of this work.

\subsection{Magnetic dimers}
\label{dimers}

\begin{figure}[htbp!]
    \begin{center}
        \includegraphics[width=0.9\columnwidth]{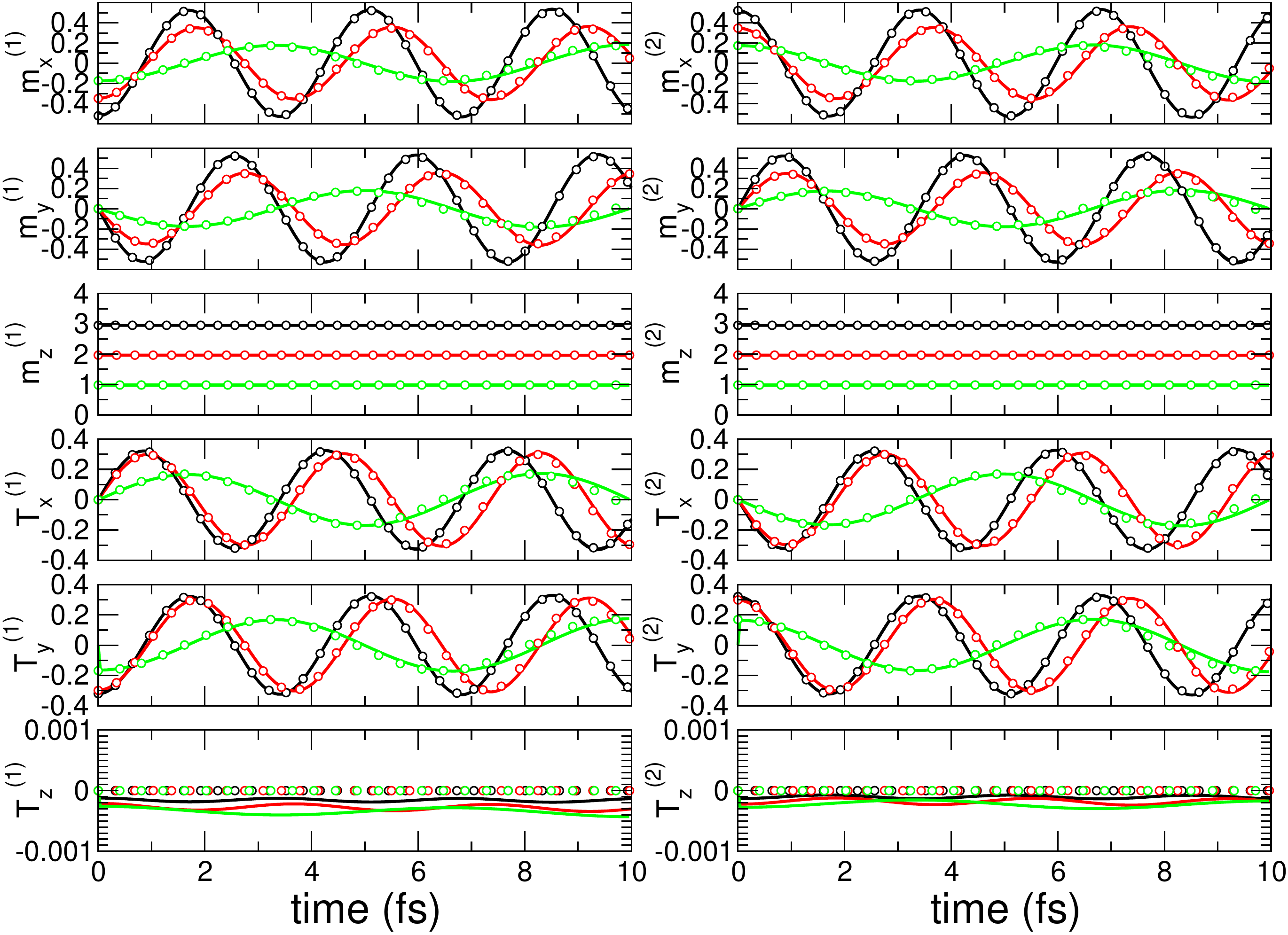}
    \end{center}
    \caption{\label{figure1}(color online) Magnetization and torque dynamics of individual moments for for dimers of Fe (black), Co (red) and Ni (green). TBSD (resp. ASD) results are in solid lines (resp. circles). Unit of torques is PHz. Initial conditions are ${\bm m}_1=g(-\sin(10^{\circ}),0,\cos(10^{\circ}))$, ${\bm m}_2=g(\sin(10^{\circ}),0,\cos(10^{\circ}))$, where $g$ are the SCF Landé factor for each atom (see Appendix~\ref{calculated_exchange}).}
\end{figure}

Many studies have already addressed the spin dynamics of both quantum and classical Heisenberg dimers~\cite{mentrupSpinDynamicsQuantum1999,*efremovHeisenbergDimerSingle2002,*kolezhukDynamicsAnisotropicSpin2004,*cabotQuantumSynchronizationDimer2019}, not always systematically by looking the temporal dynamics of each of their individual moments.
Using the method described in Sec.~\ref{spin_dynamics}, we studied the time evolution of the net magnetic moments, here treated as a classical tridimensional vectors, for magnetic dimers of Fe, Co and Ni.
First, Eq.~\eqref{precession_eq} is solved and the precession of these magnetic moments is analyzed without damping, by starting from a tilted angle of $10^{\circ}$ from the $z$-axis for each atomic site, as the initial configuration.
Then by using the method presented in the Appendix~\ref{calculated_exchange}, our findings are compared with an atomistic spin dynamics approach using the exchange coupling $J$ extracted from the angular dependence of the total energy. 
Our results, depicted in Fig.~\ref{figure1}, show that all the three dimers behave well as under the Heisenberg interaction in the studied limit, i.e. the effective field $B^{eff}_{i}$ can be described by a constant isotropic exchange, Eq.~\eqref{eq1}, that does not depend on the instantaneous magnetic configuration.
As shown in Appendix~\ref{calculated_exchange}, between $\theta=0^{\circ}$ and $\theta=10^{\circ}$ the fit between the energy calculated from the TB onto a Heisenberg Hamiltonian works perfectly, but that does not hold true for higher angles. 
It means that a simple bi-linear Heisenberg Hamiltonian is not enough to describe the system globally, but only locally with respect to the magnetic configuration.
Because the $z$-component of the magnetization is constant in time, the $z$-component of the ASD torque is exactly zero, which is not the case in the TB dynamics.
However, this can be consistently monitored by decreasing the timestep used to integrate the precession equation, Eq.~\eqref{precession_eq}.

We can monitor that the precession frequency, as calculated in the appendix~\ref{spindynamics_dimers}, is well reproduced by the TB calculations.
 
\subsection{Magnetic trimers}
\label{trimers}

It is known in the literature that in some specific situations, the exchange coupling and Dzyaloshinskii-Moriya interactions calculated from the ferromagnetic (FM) state are not a good fit for predictions of magnetic properties, e.g. close to the paramagnetic state~\cite{shallcrossInitioEffectiveHamiltonian2005,*rubanTemperatureinducedLongitudinalSpin2007} or the transition from the FM to the skyrmion phase~\cite{muhlbauerSkyrmionLatticeChiral2009}.
This is mainly because that in these scenarios, interactions of higher order play an important role and even sometimes a central role, such as the value of considering the 4-spin interaction in case of stabilizing the skyrmion phase in hexagonal Fe film of one-atomic-layer thickness on the Ir(111) surface~\cite{heinzeSpontaneousAtomicscaleMagnetic2011}.
These higher order interactions can be seen as if the exchange constants become kinetic functions of the magnetization state, a possibility theorized long time ago~\cite{chaoKineticExchangeInteraction1977,*chaoCanonicalPerturbationExpansion1978}.
One could argue that it is only needed a high-order more specific spin-Hamiltonian to describe the problem, but in some other cases the so called beyond-Heisenberg interactions can also be present, i.e. interactions that cannot be mapped into a spin-Hamiltonian~\cite{kvashninMicroscopicOriginHeisenberg2016} or cases where the Heisenberg picture is simply broken~\cite{mankovskyExtensionStandardHeisenberg2020}.
Our goal here is to explore the limits and differences between the spin dynamics features using a spin-Hamiltonian and our presented here TB spin dynamics method.

\begin{figure}[htbp!]
    \begin{center}
        \includegraphics[width=0.9\columnwidth]{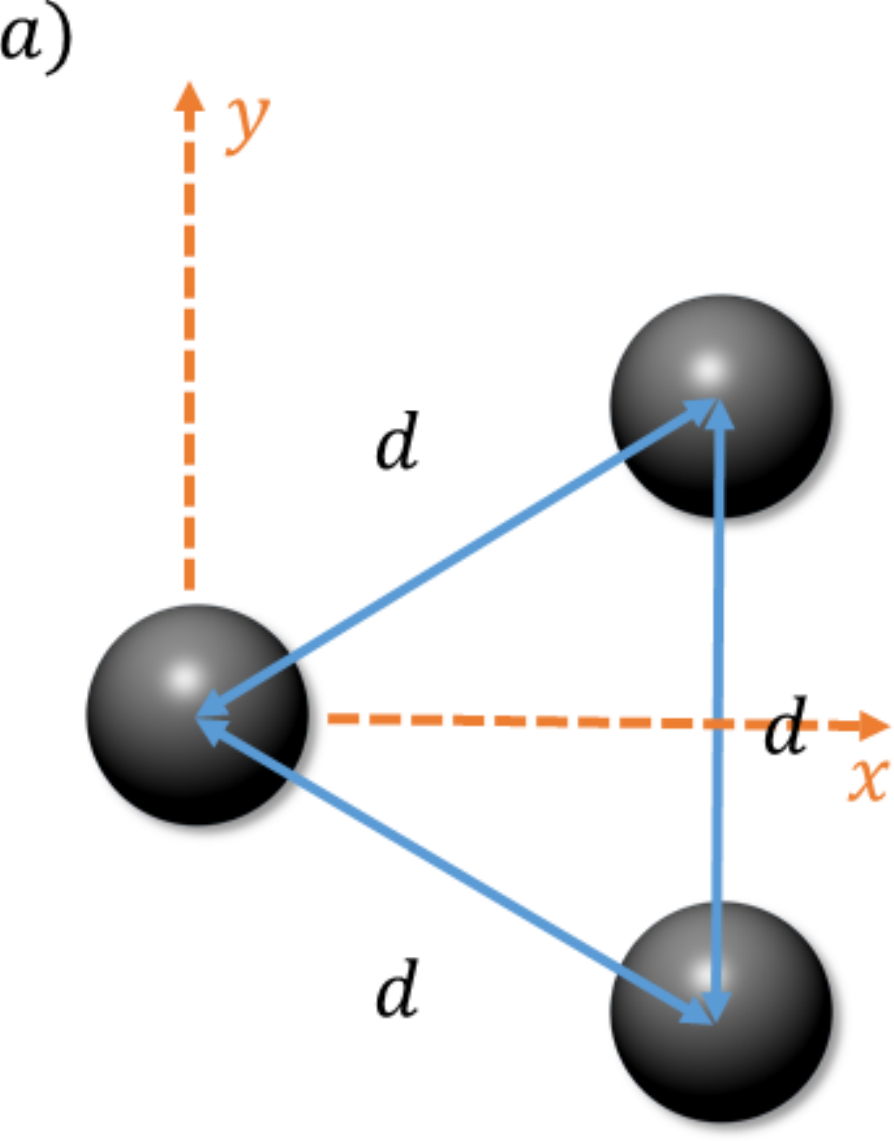}
    \end{center}
    \caption{\label{figure2}(color online) a) Schematic representation of the the equilateral triangle trimer.}
\end{figure}

\begin{figure}[htbp!]
    \begin{center}
        \includegraphics[width=0.9\columnwidth]{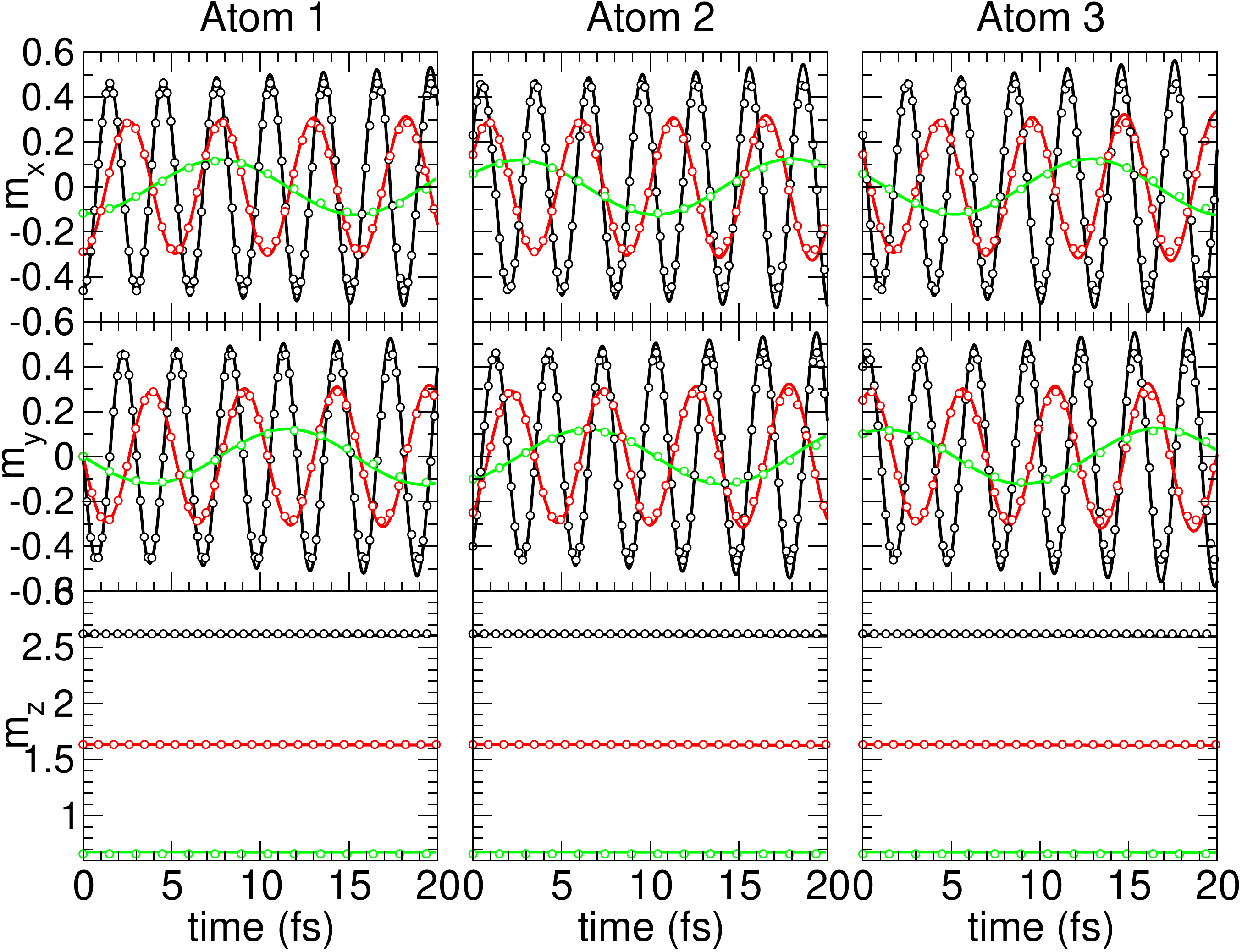}
    \end{center}
    \caption{\label{figure3}(color online) Magnetization dynamics of Fe (black), Co (red) and Ni (green) triangle trimers. TBSD (resp. ASD) results are in solid (resp. circles) lines. Initial conditions are $\bm{m}_1(0)=g(-0.17365,0.0,0.98481)$, $\bm{m}_2(0)=g(0.08682,-0.15038,0.98481)$, $\bm{m}_3(0)=g(0.08682,0.15038,0.98481)$, where $g$ are the SCF Landé factor for each atom (see text).}
\end{figure}

\begin{figure}[htbp!]
    \begin{center}
        \includegraphics[width=0.9\columnwidth]{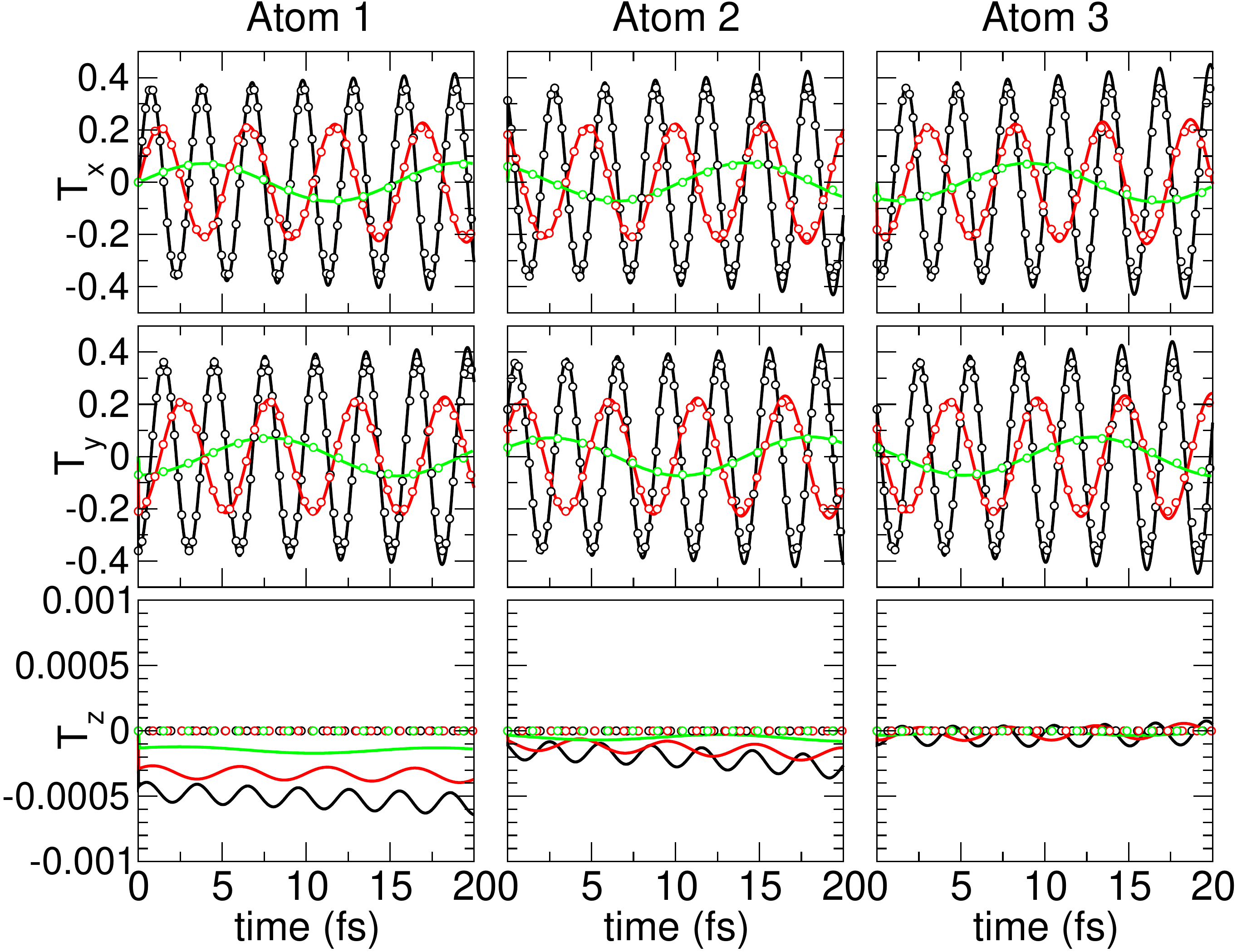}
    \end{center}
    \caption{\label{figure4}(color online) Torque dynamics of Fe (black), Co (red) and Ni (green) triangle trimers. TBSD (resp. ASD) results are in solid (resp. circles) lines. Units of the torques are in PHz. Initial conditions are identical than those in Fig.~\ref{figure3}.}
\end{figure}

In order to do that, the magnetization dynamics of magnetic equilateral triangle trimers of Fe, Co and Ni is explored, as can be seen in Fig.~\ref{figure2}
The magnetization dynamics of Fe, Co and Ni triangle trimers, are depicted in Fig.~\ref{figure3} as well as the torques in Fig.~\ref{figure4}.

In order to evaluate the exchange coupling between the magnetic moments in this case, an analogous procedure to what was done to the dimer is performed, more precisely described in the Appendix~\ref{calculated_exchange}.
Fitting with the energy obtained from the TB calculation, the parameters are reported in Appendix~\ref{calculated_exchange}.
Note that in this particular case $J_{12}=J_{23}=J_{31}$ due to the symmetry.
Initially, self-consistent calculations under the angular penalty function were performed in order to determine the magnetic moments of each atom in the system.
With that information, one performs simulations of the magnetization dynamics using the spin Hamiltonian, Eq.~(\ref{eq1}).
Parallel to it, the process described in Sec.~\ref{spin_dynamics} is followed, the magnetization dynamics is calculated and the comparison between the different methods is shown in Fig.~\ref{figure3}.
Similarly to the dimers case, the systems here presented show themselves as Heisenberg systems within the studied limit, e.g. $\theta = 10^{\circ}$, when calculating the precession of the magnetic moments around the z-axis.

So far, these limits have served to prove the reliability of our method, and not to justify the extra computational cost introduced to reproduce the behavior of an ASD approach.
In the next section we exhibit the simplest situation that demonstrate its relevance.

\subsection{Configuration dependence of the exchange coupling parameters $J_{ij}$}
\label{configdependence}

The task of finding a reliable Hamiltonian to describe variations of magnetic configurations is not straightforward. 
Continuous efforts have been made throughout the years in the attempt to understand the microscopic origin of these exchange parameters and their consequences~\cite{adlerHeisenbergFerromagnetHigherorder1979}. 
Recently, a method to calculate the exchange coupling parameter $J_{ij}$ for any given magnetic configuration, via first-principles simulations, was developed and applied to study these interactions on Fe-bcc~\cite{szilvaInteratomicExchangeInteractions2013}.
In fact, these configuration dependent $J_{ij}$'s significantly improved the spin-wave dispersion comparison between the theory and the experiment.
Within the TB approximation, Ref.~\cite{streibExchangeConstantsLocal2021} reports a configuration dependence of the exchange parameters by comparing various effective field $B_{eff}$ between the Heisenberg model and direct TB calculations.
Moreover, it is crucial to understand the relevance of higher order parameters in the expansion of the magnetic Hamiltonian, e.g. and bi-quadratic terms, 3-spins, 4-spins, etc., as can be seen in works like Refs.~\cite{heinzeSpontaneousAtomicscaleMagnetic2011,brinkerChiralBiquadraticPair2019} and \cite{diasProperImproperChiral2021}. 
Lastly Ref.~\cite{drautzSpinclusterExpansionParametrization2004} as implemented in Ref.~\cite{szunyoghAtomisticSpinModel2011}, offers an attractive solution to the problem of a statistically under-represented magnetic reference state, but at a cost of a span of the entire magnetic configuration space.
In principle, this allows the derivation of effective exchange coupling constants that average the effect of more than 2 independent configurations of spins.
Unfortunately this statistical method is more suitable in the dilute magnetic limit and appears not adequate to capture the magnetic behavior of a single specific dimer or trimer.
Moreover its implementation for alloys is complex.

\begin{figure}[htb!]
    \begin{center}
        \includegraphics[width=0.9\columnwidth]{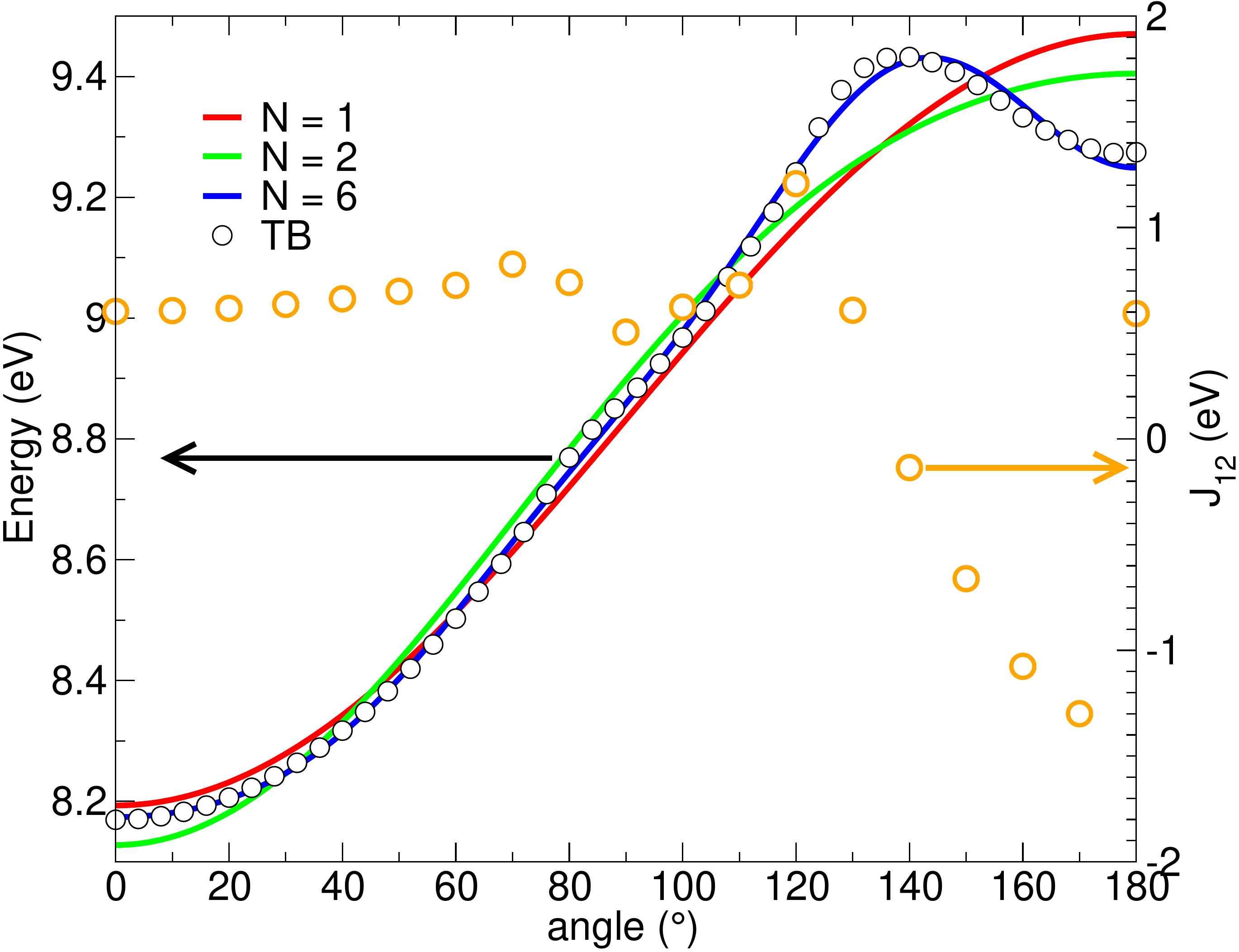}
    \end{center}
    \caption{\label{figure5}(color online) TB total energy as a function of the angle between the two magnetic moments of an Fe dimer on the left y-axis and the $N=1$ exchange coupling parameter derived locally for each angle, on the right y-axis. In addition, the TB total energy is globally fitted by expansion in Legendre polynomials in terms of $\cos(\theta)$. Here, $N=1$ would be the bi-linear Heisenberg Hamiltonian, $N=2$ includes the bi-quadratic term and so on so forth.}
\end{figure}

So far, we have calculated the exchange coupling parameters by fitting the energy from the TB calculations around the ground state, i.e. FM for Fe, Co and Ni. 
These past studies have revealed the non-Heisenberg behavior of Fe in particular and in order to illustrate our argument, we picked up the Fe dimer as an example.
For a dimer, one can express the total TB energy as an expansion on a basis of Legendre polynomials up to a given order $N$, such as
\begin{align}
    E(\theta)-E(0)&=\sum_{n=1}^N J_{12}^{(n)}P_n(\cos(\theta)).
\end{align}
When this series ends to $N=1$, $J_{12}^{(1)}$ is just the usual intensity of the Heisenberg coupling constant.
If this series ends to $N\ge2$, we can interpret $J_{12}^{(2)}$ as a biquadratic component of the intensity of the magnetic coupling, characterized by a beyond-Heisenberg behavior.
In the Fig.~\ref{figure5} we show on the left, the total energy of Fe dimer as a function of the angle $\theta$ between the magnetic moments of each Fe atom, along with the exchange coupling $J_{ij}^{(1)}$ calculated by fitting the Heisenberg model around the local $\theta$ (at every step of $\theta=10^{\circ}$), on the right.

It is clear from the total energy calculations that, for that case, it cannot be fitted by a simple bi-linear Heisenberg model.
We tried then to add a bi-quadratic correction to the model as $\cos^2(\theta)$, as done in Ref.~\cite{costaGroundStateMagnetic2005}, by analyzing the $P_2(\cos(\theta))=\frac{1}{2}\left(3\cos^2(\theta)-1\right)$ part of the Legendre expansion, and then reported in the Fig.~\ref{figure5} along the $N=2$ curve.
One can note that this $N=2$ term improves globally the model curve, but quite not match the TB calculations rigorously, in particular in the range of angles when the FM order is not the preferred magnetic ground state.
It is needed to go up to the 6-th order to get a reasonable fit that captures all the energetic features, including the reversal in the sign of the energy behavior at intermediate angles.
It is noteworthy to mention that the magnetic moment of each of the Fe atoms changes throughout the rotation of about 40\% (data not shown), from 3 $\mu_{B}$ (FM configuration) to $\approx 1.8 \mu_{B}$ (AF configuration); a feature that is also not covered by the Heisenberg model.
The parametric derivation of such a simple configuration space indicates the magnitude of the task at hand in much more complex systems, such as alloys and materials with non-collinear magnetic configurations as ground state.
However we argue that properties strongly dependent on small variations around the ground state, such as spin-wave spectra, are well described with a local Heisenberg Hamiltonian, as already anticipated by Holstein and Primakoff~\cite{holsteinFieldDependenceIntrinsic1940}, but we need a more precise electronic structure behavior, in order to compute the correct effective field far from the ground state and not necessarily represented by the magnon state of lowest energy.
In that scenario the effective field directly derived from the electronic structure, produces the correct dynamics in time for any directions of any local magnetic moments, without prior knowledge of any exchange values and represents, by construction, a direct solution to avoid such issue.
\section{Conclusion}
\label{Conclusion}

In this paper, we have presented a method that offers an alternative between full \textit{ab initio} and spin-Hamiltonian based spin-dynamics.
Our approach uses a penalty functional on the magnetic moments of each site in order to calculate self-consistently, at every time step, the respective effective magnetic field.
We have solved the precession equation on each site, without damping, for dimers and trimers of Fe, Co and Ni, and compared our findings with an ASD approach, where the magnetic effective field is not calculated directly from the electronic structure, but from a parameterized spin-Hamiltonian. 
The exchange coupling interaction $J$, as a parameter, was calculated by fitting the TB total energy with a parameterized spin-Hamiltonian for a range of directions of the atomic magnetic moments. 
Our results showed that within this limit, they can be seen as good Heisenberg systems locally and the comparison between the TB and ASD are fairly good. 
That is not the case where the same set of magnetic moments connect different magnetic extrema, meaning that different parametric local representations have to be calculated, which breaks the whole Heisenberg picture.
For those systems, one cannot map globally the electronic structure onto a single Heisenberg model, although these parameters still can predict with good accuracy properties of their local ground states. 
We have illustrated this situation by studying the dependance of the total energy of an Fe dimer, as a function of the angle between the atomic magnetic moments, and proved that this cannot be mapped globally into a bi-linear Heisenberg Hamiltonian only.
In fact, a high-order expansion in power of the angular directions between the atomic magnetic moments is mandatory to match the landscape of the TB energy adequately.
Finally, the TBSD here presented is a satisfying solution, with a reasonable computational cost, to study the spin-dynamics of systems that are not dominated by the pair Heisenberg's interaction only, because the construction of the \textit{ab initio} effective field is free from such hypothesis. 
This technique may serve also to investigate the dynamics of more complex magnetic systems that include spin-orbit mediated interactions in low dimensional symmetries, and appears to be both versatile and general.

\section*{Acknowledgments}
\label{Acknowledgments}

We gratefully thank to the Programme Transversal de Competences for financial support with the project DYNAMOL.

\section*{Data availability statement}
\label{DAS}

The data that support the findings of this study are available from the corresponding author, upon reasonable request.
\appendix

\section{Fixed spin moment}
\label{fixed_spin}

The fixed spin moment calculation is probably the most straightforward method, but is limited to the case of collinear magnetism and is independent of the site index.
This is to impose exactly a total magnetization of the system and therefore the total number of $\uparrow$ and $\downarrow$ electrons.
One therefore needs to define two separate Fermi levels $E_F^{\sigma}$. 
For a homogeneous system where each atom carries the same charge and the same magnetization, the total energy is 
\begin{equation}
    E_{\text{tot}}=\sum_{\alpha}^{ |\varepsilon_{\alpha}^{\uparrow}|<E_F^{\uparrow} } \varepsilon_{\alpha}^{\uparrow} +\sum_{\alpha}^{ |\varepsilon_{\alpha}^{\downarrow}|<E_F^{\downarrow} } \varepsilon_{\alpha}^{\downarrow}
    + \frac{1}{4} I m^2,
\end{equation}
where $\varepsilon_{\alpha}^{\sigma}=\varepsilon_{\alpha}^{0}-\frac{1}{2}Im \sigma$.
Then the total energy can be rewritten as
\begin{equation}
   E_{\text{tot}}=\sum_{\alpha}^{ |\varepsilon_{\alpha}^{0}|<E_F^{\uparrow}+\frac{1}{2}I m } \varepsilon_{\alpha}^{0} +\sum_{\alpha}^{ |\varepsilon_{\alpha}^{0}|<E_F^{\downarrow}-\frac{1}{2}I m } \varepsilon_{\alpha}^{0}
    -\frac{1}{4} I m^2.
    \label{E_FSM}
\end{equation}
Consequently, the derivative of the total energy with the magnetization becomes simply proportional to the difference of Fermi's energies
\begin{equation}
    \frac{dE_{\text{tot}}}{dm}=\frac{(E_F^{\uparrow}-E_F^{\downarrow})}{2}.
\end{equation}
An effective field $B^{\text{eff}}=-(E_F^{\uparrow}-E_F^{\downarrow})/2$, aligned to these moments, can be defined.
It comes out that at the extrema of $E_{\text{tot}}$, the two Fermi levels are equal and the effective field becomes zero.
By looking at the sign of the second derivative of the energy around $m=0$, this is simple to recover the Stoner criterion as described in the reference~\cite{blundellMagnetismCondensedMatter2014}.
Although useful, the fixed spin moment method is limited to rather homogeneous systems.

\section{Penalty method for atomic spin moment}
\label{pen_fixed_spin}

Let us consider the case where a given magnetization $\bm{m}_i^{\text pen}$ is imposed on each atom.
A quadratic penalty term as $E^{\text pen}_i=\frac{\lambda}{2}\norm{\bm{m}_i-\bm{m}_i^{\text pen}}^2$ can be added to each site, where $\lambda$ is a large positive number.
In principle $\lambda$ should go to infinity, but in practice a good compromise is found by increasing its value and to check the convergence of the desired quantity computed with.
However, this problem can be circumvented by implementing the Augmented Lagrangian Method, that introduces a quadratic constraint term in the renormalized Hamiltonian, such as the $\lambda$ parameter remains finite~\cite{liEfficientAugmentedLagrangian2013}.
This is at the cost of an additional computational complexity and the penalization approach with a sufficient large $\lambda$ term is preferred. 

This consists to supplement Eq.\eqref{Hvec} with the term $\lambda({\bm m}_i-{\bm{m}_i^0})\ket{i}\bra{i}$. 
Consequently, the on-site diagonal renormalization term can formally be written $U_i \Delta n_i \sigma_0 -( \bm{B}^{\text Stoner}_i+\bm{B}^{\text pen}_i)\cdot\bm{\sigma} $ with $\bm{B}^{\text Stoner}_i=\frac{1}{2} I_i \bm{m}_i$,  $\bm{B}^{\text pen}_i=-\lambda(\bm{m}_i-\bm{m}_i^{\text pen})$ and $\Delta n_i=(n_i-n_i^0)$.
The total energy should be corrected accordingly by the double counting terms and reads
\begin{align}
    E_{\text{tot}}[\{\bm{m}_i^0\}]&=\sum_{ \alpha }^{\text{occ}} \varepsilon_{\alpha}
    -\frac{1}{2}\sum_i U_i ((n_i)^2-(n_i^0)^2) \nonumber\\
    &+\frac{1}{4}\sum_i I_i \norm{\bm{m}_i}^2
    -\frac{\lambda}{2} \sum_i  (\norm{\bm{m}_i}^2-\norm{\bm{m}_i^{\text pen}}^2).
    \label{eq:double-counting}
\end{align}
In the limit $\lambda \to \infty$, $-\lambda(\bm{m}_i-\bm{m}_i^{\text pen})\approx\bm{B}^{{\text pen} \infty}_i$ and $\bm{m}_i \approx \bm{m}_i^{\text pen}$. 
Consequently, the corresponding double counting term $-\frac{\lambda}{2} (\norm{\bm{m}_i}^2-\norm{\bm{m}_i^{\text pen}}^2)$ can be rewritten as $\bm{B}^{{\text pen}\infty}_i \cdot \bm{m}_i^{\text pen}$.

The fixed spin moment can be seen as a special case of the penalty method applied for collinear magnetism with only one type of atom. 
The term $-B^{\text pen} \sigma$ in the renormalized Hamiltonian just shifts rigidly the eigenvalues by $-B^{\text pen}$ for $\uparrow$ spin and $B^{\text pen}$ for $\downarrow$ spin, such as $\varepsilon_{\alpha}=\varepsilon_{\alpha}^0-\frac{1}{2} I m \sigma-B^{\text pen}\sigma$.
The total energy of Eq.\eqref{E_FSM} is recovered once provided $E_F^{\sigma}=E_F+\sigma B^{\text pen}$.
Then one gets $B^{\text pen}=\frac{1}{2}(E_F^{\uparrow}-E_F^{\downarrow})=-B^{\text{eff}}$.

\section{Solution of the spin dynamics of ferromagnetic dimers}
\label{spindynamics_dimers}

The motion of each individual moments of ferromagnetic dimers within the Heisenberg interaction is a two-body problem admitting an exact solution.
Let's $\Omega_s^0\equiv J^0/\hbar$ the magnitude of the exchange pulsation and $E=-J^0\bm{m}_1\cdot\bm{m}_2$ its interaction energy, with $J^0>0$.
The motion of each undamped moment is the solution of a set of 2 coupled equations of precession, which are
\begin{equation}
    \label{sddimer}
    \begin{aligned}
        \frac{d{\bm m}_1}{dt}&=\Omega_s^0\bm{m}_2\times{\bm m}_1,\\
        \frac{d{\bm m}_2}{dt}&=\Omega_s^0\bm{m}_1\times{\bm m}_2,
    \end{aligned}
\end{equation}
with the given initial conditions ${\bm m}_1(0)$ and ${\bm m}_2(0)$.

Equivalently when using an Heisenberg Hamiltonian with normalized vectors $E=-J\bm{e}_1\cdot\bm{e}_2$, with $J=J^0m^2$ (where $m$ is the amplitude of the magnetization) we get the coupled evolution equations:

\begin{equation}
    \label{sddimer2}
    \begin{aligned}
        \frac{d{\bm e}_1}{dt}&=\Omega_s\bm{e}_2\times{\bm e}_1,\\
        \frac{d{\bm e}_2}{dt}&=\Omega_s\bm{e}_1\times{\bm e}_2,
    \end{aligned}
\end{equation}
with  $\Omega_s\equiv J/\hbar$.
This motion is decoupled in the frame of the magnetization ${\bm{e}}\equiv\left({\bm{e}_1}+{\bm{e}_2}\right)$.
In this frame, by combining Eqs.\eqref{sddimer2} together, one finds $\frac{d{\bm e}}{dt}=\bm{0}$ and consequently ${\bm e}$ is a constant vector given by the initial conditions ${\bm e}=\left({\bm{e}_1}(0)+{\bm{e}_2(0)}\right)$.
By noting that $\Omega_s\bm{e}_2\times{\bm e}_1=\Omega_s(\bm{e}_1+\bm{e}_2)\times{\bm e}_1=\Omega_s\bm{e}\times{\bm e}_1$,
Eqs.~\eqref{sddimer2} become fully decoupled:
\begin{equation}
    \label{MLdimer}
    \begin{aligned}
        \frac{d{\bm e}_1}{dt}&=\Omega_s{\bm e}\times{\bm e}_1,\\
        \frac{d{\bm e}_2}{dt}&=\Omega_s{\bm e}\times{\bm e}_2.
    \end{aligned}
\end{equation}
Then the motion of each of these unit vectors ${\bm e}_i$ is simply the motion of a vector in a constant field.
Its solution is given by the Rodrigues' formula~\cite{rodriguesLoisGeometriquesQui1840,*thibaudeauThermostattingAtomicSpin2011} 
\begin{align}
    {\bm e}_i(t)&=\cos(\Omega_s t){\bm e}_i(0)+\sin(\Omega_s t)\bm{e}+(1-\cos(\Omega_s t))\chi_i{\bm e}_i(0)\times \bm{e},
    \label{rodrigues}
\end{align}
where $\chi_i\equiv \bm{ e}_i(0)\cdot \bm{e}$.

The same reasoning can be derived for trimers of identical atoms with the same exchange parameters applied up to the first neighboring shell, in between. 
In that very specific case, each atomic spin follows the same equation of precession, namely
\begin{equation}
    \frac{d{\bm e}_i}{dt}={\Omega}_s{\bm e} \times {\bm e}_i,
\end{equation}
with ${\bm{e}}\equiv\sum_{i=1}^3{\bm e}_i(0)$, where ${\bm e}$ is found to be constant of motion.
Consequently for trimers with identical atoms and interactions, the precession frequency, and thus the value of the exchange parameter, can be measured from a single motion of any spins, as depicted in Figs.~\ref{figure3} and \ref{figure4}.     

\section{Calculation of the exchange coupling parameters}
\label{calculated_exchange}

The macroscopic nature of the exchange coupling parameters and how they are influenced by the various circumstances have been widely discussed in the literature.
The Bethe-Slater~\cite{slaterAtomicShieldingConstants1930,*slaterCohesionMonovalentMetals1930,*chikazumiPhysicsFerromagnetism1997} (BS) curve explains in an insightful way, by means of direct exchange and the distance between nearest-neighbor (NN) atoms, the trends followed by ferromagnetism (FM) and antiferromagnetism (AFM) ground state of the 3d transition metals from bcc Cr to hcp Co.
Recent studies~\cite{cardiasBetheSlaterCurveRevisited2017} have shown that, even for the bulk case of such elements, the BS curve reveals a complicated background behind the macroscopic picture.
Such NN interactions depend not only on the distance but also the symmetry and their bonds, i.e. influenced by the crystal field.
That kind of dependence has also been seen in supported nanoclusters~\cite{mavropoulosExchangeCouplingTransitionmetal2010,*rodriguesFirstprinciplesTheoryElectronic2016,*belabbesHundRuleDrivenDzyaloshinskiiMoriya2016}, where for the same distance, different values for the exchange coupling parameter can be found.
In case of small clusters, like the dimers and trimers studied here, the local density of states of each atom is very localized, which set apart the majority band from the minority band.
It implies in a large band splitting that directly affects the value of the of the exchange coupling parameter~\cite{bergmanMagneticInteractionsMn2006,frota-pessoaExchangeCouplingTransitionmetal2000}.
As coordination number increases, the hybridization results in the broadening of such bands, shifting the center of it closer to the Fermi energy, thus decreasing the value of the exchange coupling parameter as the coordination number increases~\cite{bezerra-netoComplexMagneticStructure2013,igarashiMagneticPropertiesFexCo1xnanochains2014}.
Moreover, the results here presented follow this logic, as well as the BS curve trend.

\begin{figure}[htb!]
    \begin{center}
        \begin{tabular}{cc}
            \includegraphics[width=0.45\columnwidth]{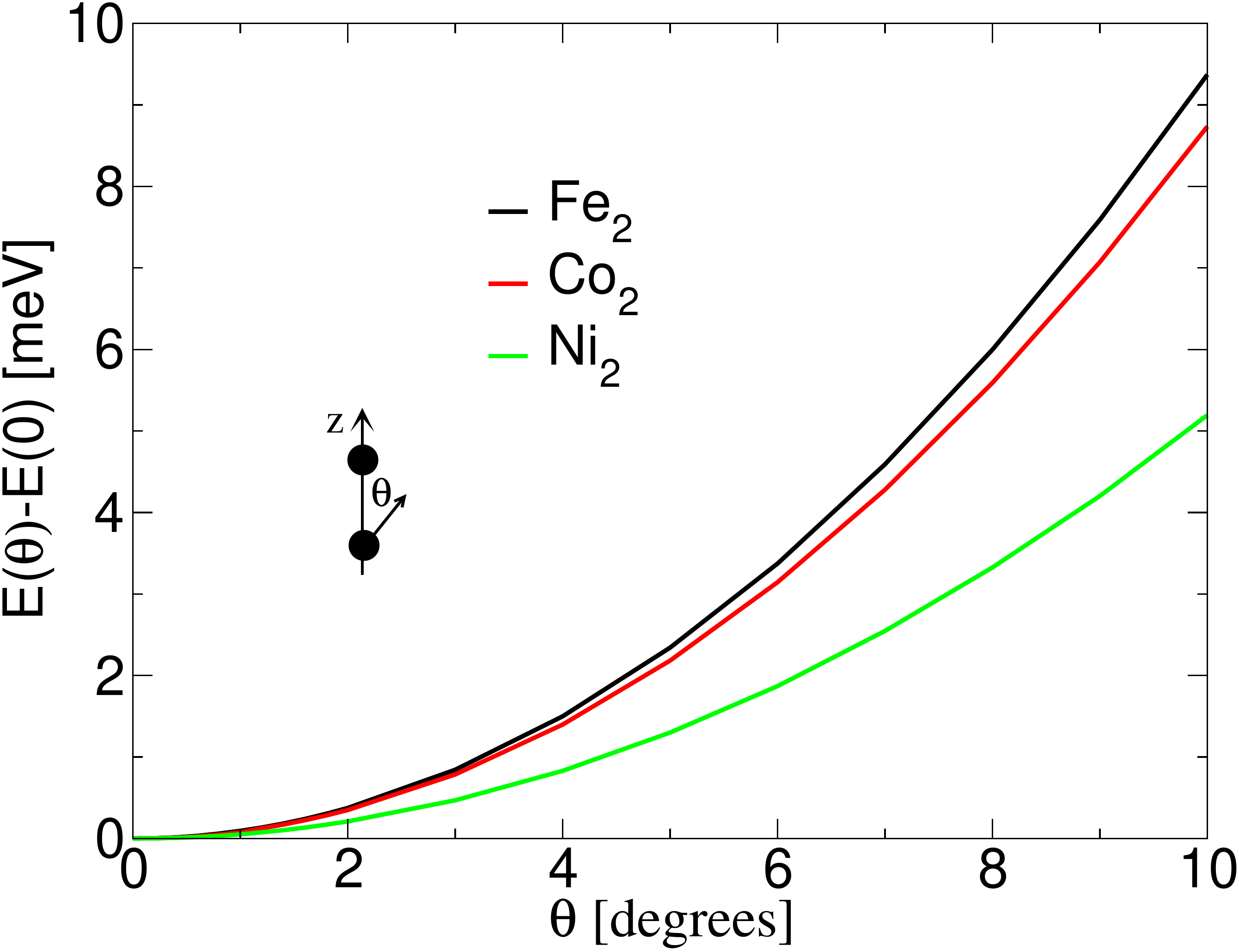}&\includegraphics[width=0.45\columnwidth]{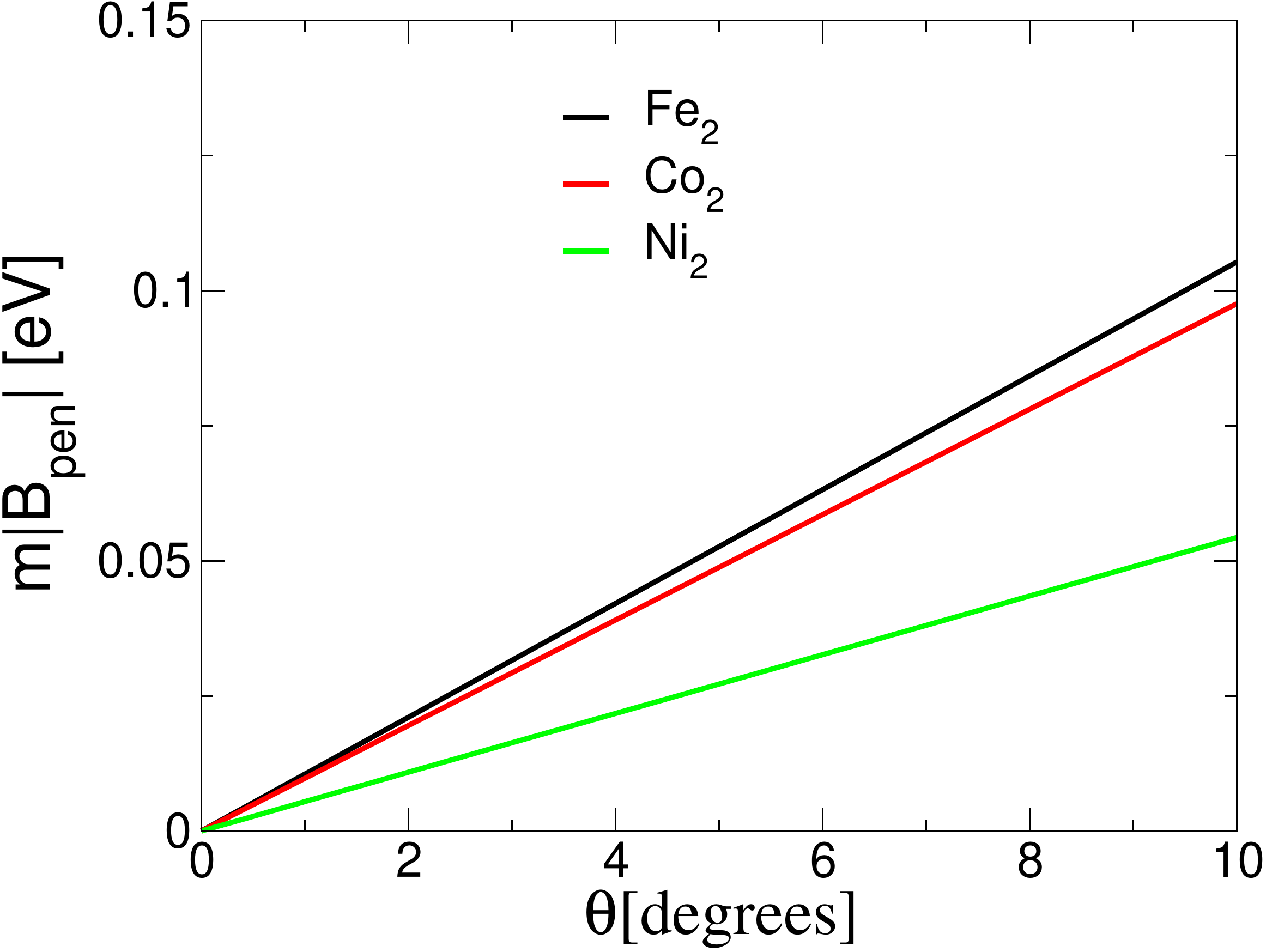}\\
            \includegraphics[width=0.45\columnwidth]{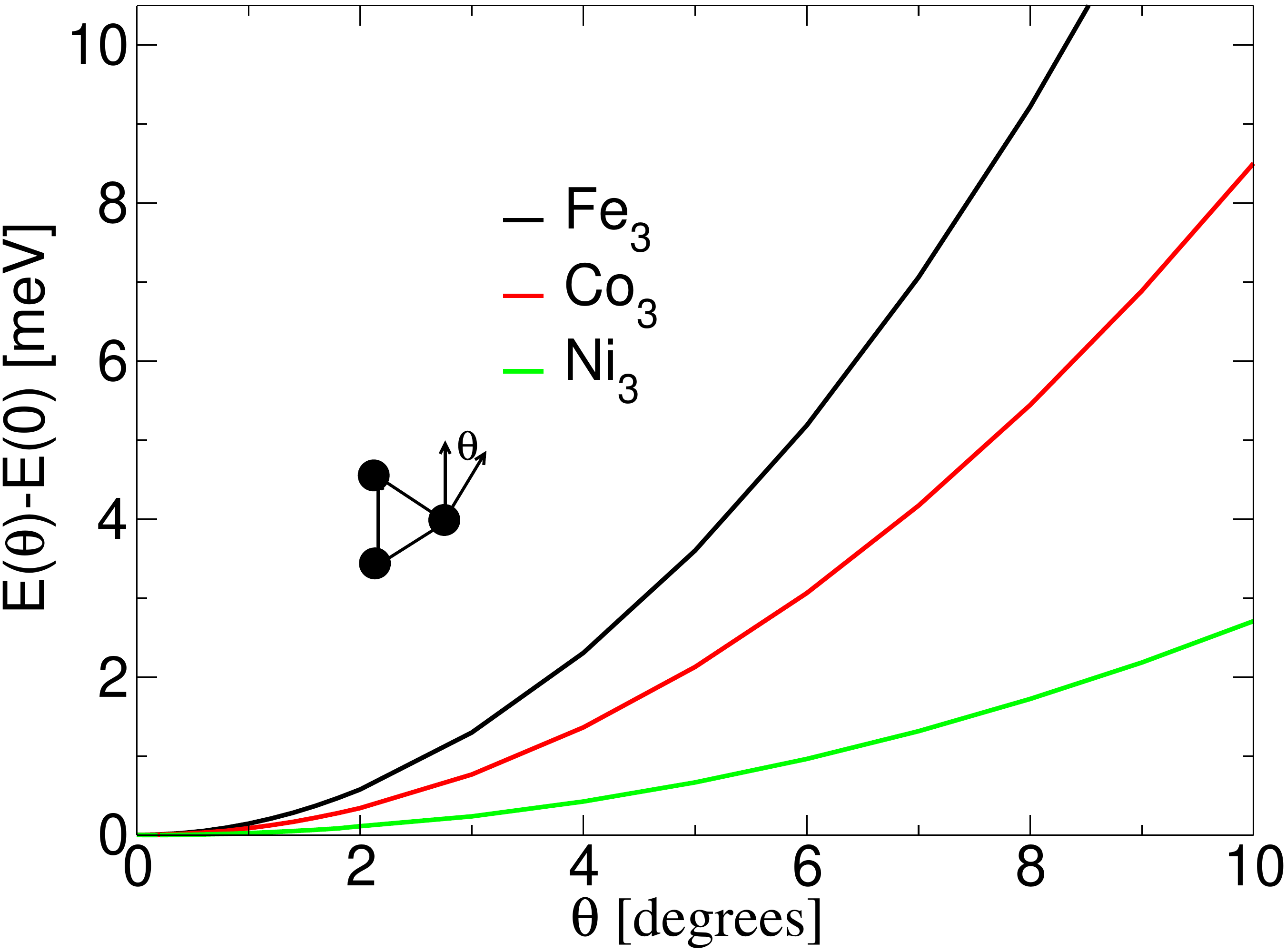}&\includegraphics[width=0.45\columnwidth]{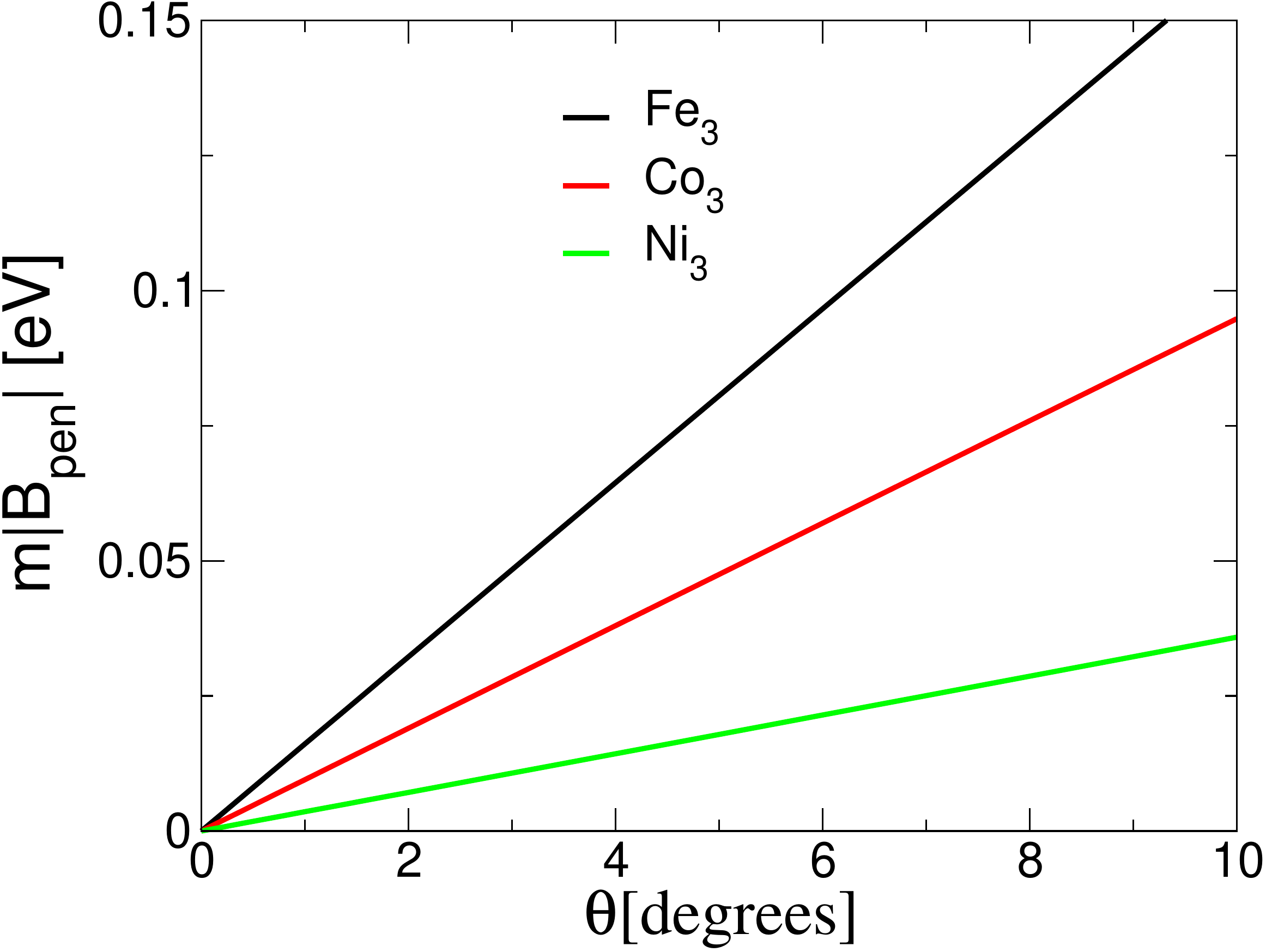}
        \end{tabular}
    \end{center}
    \caption{\label{figure6} (color online) Calculated total energy (left) and effective field (right) from the tight-binding model for dimers and trimers of Fe, Co and Ni (interatomic distance of 2\AA), in black, red and green, respectively. Dimers on top and trimers on the bottom. The corresponding fits from an Heisenberg Hamiltonian fall exactly on top of the tight-binding curves (and are not shown), suggesting that the Heisenberg's pair interaction dominates these systems within that range of $\theta$.}
\end{figure}

For each of the magnetic configurations, the total energy is computed with the TB parameters found in reference~\cite{barreteauEfficientMagneticTightbinding2016}.  
When only one rotating single magnetic moment is considered, the total energy in the Heisenberg model can be written as a function of the angle with the $z$-axis, labelled $\theta$.
For the dimer it reads
\begin{equation}
    E_{\text{dimer}}(\theta)-E_{\text{dimer}}(0)=J_{\text{dimer}}(1-\cos(\theta)),
    \label{dimer-fit}
\end{equation}
and for the trimer
\begin{equation}
    E_{\text{trimer}}(\theta)-E_{\text{trimer}}(0)=2J_{\text{trimer}}(1-\cos(\theta)).
    \label{trimer-fit}
\end{equation}
As seen in Fig.~\ref{figure6}, Eqs.~\eqref{dimer-fit} and \eqref{trimer-fit} can be fitted with the total energy computed in the TB approximation, in order to find the respective exchange coupling parameters $J$.
For the dimer, it is obvious that $J_{12}=J_{21}\equiv J_{\text{dimer}}$ and for the trimer, because of the $C_3$ symmetry, $J_{12}=J_{23}=J_{31}\equiv J_{\text{trimer}}$ also.
The fact that the fitting and the energy curve fall on top of each other, means that both $J_{\text{dimer}}$ and $J_{\text{trimer}}$ are constants within the limit considered of $\theta$, i.e. the electronic interaction in these systems is dominated mainly by the Heisenberg's pair interaction~\eqref{eq1} in that range.
The computed values taken for an equal distance $d=2\text{\AA{}}$ between atoms are reported in the tables~\ref{Jvaluesfordimers} and \ref{Jvaluesfortrimers}.

Finally another strategy has been tested to evaluate the exchange parameters. 
Instead of considering the total energy variations $E(\theta)$ as the reference quantity, we have fitted the variation of the effective field ${\bm B}^{\text{pen}}$ as a function of the deviation angle $\theta$. 
Indeed it is straightforward to show that $\norm{\bm{B}^{\text{pen}}}\norm{\bm{m}}$ is equal to $J\sin\theta$ for the dimer and $2J\sin \theta$ for the trimer, respectively.
The results are reported in parenthesis in the tables ~\ref{Jvaluesfordimers} and \ref{Jvaluesfortrimers}.
The agreement between the two approaches is good and could be systematically improved by increasing the penalization constant $\lambda$.

\begin{table}
    \begin{tabular}{l|c|c}
        & g ($\mu_B$) & $J_{\text{dimer}}$ (eV)\\
        \hline
        Fe&3&0.616 (0.605)\\
        Co&2&0.574 (0.561)\\
        Ni&1&0.341 (0.312)\\
    \end{tabular}
    \caption{\label{Jvaluesfordimers}Values of the computed SCF magnetization and exchange parameter for dimers (interatomic distance of 2\AA) calculated in the TB approximation. In parenthesis is shown the result obtained from the fit of the effective field.}
\end{table} 
\begin{table}
    \begin{tabular}{l|c|c}
        & g ($\mu_B$) & $J_{\text{trimer}}$ (eV)\\
        \hline
        Fe&2.6666&0.442 (0.463)\\
        Co&1.6666&0.279 (0.273)\\
        Ni&0.6666&0.089 (0.103)
    \end{tabular}
    \caption{\label{Jvaluesfortrimers}Values of the computed SCF magnetization and exchange parameter for equilateral triangle trimers calculated in the TB approximation. In parenthesis is shown the result obtained from the fit of the effective field.}
\end{table}

\pagebreak
\bibliographystyle{apsrev4-2}

\end{document}